# Public-Private Cooperation in the Port of Rotterdam: An Agent-Based Toy Model of Cooperative Efforts Against Cocaine Trafficking

Frederike Oetker and Charlotte Geertman

**Abstract**

Cocaine trafficking poses a persistent governance challenge in the Port of Rotterdam (PoR), requiring cooperation between public and private actors. This paper develops an Agent-Based toy Model (ABM) coupled with a Causal Loop Diagram (CLD) to explore interactions between trust, cooperation, competition, pressure, enforcement, and actor skills in public-private cooperation (PPC). Based on empirical data of PPC in the PoR, we explore responses to parameter shocks across different initial configurations. In this model, effectiveness of PPC is strongly state-dependent and measures institutional, relational, behavioral, and resource readiness. The toy model demonstrates that isolated, single-track measures yield limited long-term gains modelled over 5 years due to structural system inertia, whereas coordinated multi-parameter strategies produce substantial improvements. Specifically, combining trust-building, cooperation, and moderate pressure with reduced regulatory enforcement friction raises effectiveness by 25% from a baseline effectiveness, compared to just 13% for an isolated increase in cooperation. Acknowledging the exploratory nature of the model, these findings offer exploratory hypotheses on port security governance. Future work should validate the model against test cases and additional expert input.

## 1 Introduction

Cocaine trafficking is a serious security concern affecting many ports, including Europe's largest: the Port of Rotterdam (PoR) (Sergi et al. 2021; Staring et al. 2019). The issue endangers the safety of port employees, harms the reputation of the port, and threatens the resilience of society at large (Staring et al. 2019). This places public and private actors under great public and political pressure to respond. These actors include transport companies, terminals, Customs, Seaport Police and others (Staring et al. 2019). Actors try to bring together their differing interests and resources in this high-pressure environment, which has resulted in public-private cooperation (PPC) in the PoR against cocaine trafficking (Staring et al. 2019). Cooperative efforts include formalized partnerships, information-sharing strategies, and shared initiatives to boost organizational awareness and resilience (Bisschop et al. 2022; Staring et al. 2019). The level of PPC established in the PoR is considered unique, given the port's size, the severity of the cocaine trafficking issue, and the stakes involved regarding business continuity and public security (Geertman 2026). Developing and consolidating this cooperative approach is challenging, a topic that is widely discussed in criminological and governance literature (Nøkleberg 2020; Quéro and Dupont 2019; Waardenburg et al. 2016). Challenges include agreeing on intervention scope and building trust between organizations, while respecting the core business of all organizations.

To explore to possibility of adressing these governance challenges from a quantitative perspective and to add to the criminological and governance debate, we present a multilayer Agent-Based toy Model (ABM) coupled with a CLD based on a case study of the PoR. Our toy model simulates behaviors, skills, and relational parameters of public and private actors in their cooperative efforts. Our goal is to improve our understanding of these dynamics and to experiment with them, aiming to identify the hypothetical optimal conditions for the modeled capacity of the PPC system to function effectively, as reflected in actor skills and relationship parameters.

### 1.0.1 Actor Skills and Relationship Parameters

This study is based on interviews with 14 actors gathered by one of the authors. This is a criminology and governance researcher referred to in this study as domain expert. These interviews were carried out in light of a recent study on the development of PPC against cocaine trafficking in the PoR between 2004 and 2024 (Geertman 2026). To protect the privacy of the actors, organizational details are anonymized. The network contains 14 actors and 72 ties. The network density is 0.79, with an average clustering of 0.8 and an average degree of 10.3. The average tie strength is 0.16, and the most central actor is Public Actor 1. See in the Supplementary Appendix Figure 2 for an overview of the actor network. See Figure 4 for a detailed breakdown of each actor. Actor skills range between low, medium and high per skill.

We build on the FREIDA methodology to couple a CLD and an ABM, please refer to the FREIDA methodology for a detailed overview (Oetker et al. 2026). This methodology enables qualitative and quantitative data inputs to become the basis of a validated ABM, using methodologies such as ODD+D for data collection. We followed the FREIDA model building process consisting of Knowledge and Data Acquisition (interpreting interview transcripts to identify and contextualize the skills and relational mechanisms of the actors, while the existing literature on criminology and governance provides the conceptual basis for their definitions), Model Development and using operationalization (the empirical material informing which mechanisms were represented and their direction of influence, rather than being used to statistically estimate model parameters) arriving at a calibrated computational toy model. Where the empirical material did not provide sufficient information to quantify effect sizes, update rates, or decay mechanisms, these values were specified as modeling assumptions based on domain expertise and theoretical reasoning in collaboration between the authors. Accordingly, this is a mechanism-exploration model rather than a predictive governance model: its purpose is to examine how combinations of governance conditions, actor interactions, feedback mechanisms, and external shocks may generate different system-level effects, rather than to predict governance efforts. The parameter origins are documented in the Supplementary Appendix in Table 5. The parameters were narrowed down to the five core ones in multiple Group Model Building sessions with the domain expert and criminological and governance literature. Additionally, a workshop was held with a panel of 10 interdisciplinary researchers, ranging from criminology to serious games and computational science, to discuss the skill and parameter values and rules.

Using an ABM or CLD to study governance efforts has been done before, see for example the works of Balke et al. (Balke et al. 2013), Canales et al. (Canales et al. 2026) and Crabolu et al. (Crabolu et al. 2023). However, to align this ABM and CLD with our focus on PPC in port security specifically, our toy model builds on a theoretical framework used in criminological literature rather than governance literature. This framework, called *nodal governance*, posits that security governance is achieved through an interplay between public and private nodes (i.e. actors) within a network, with each node contributing specific skills (Burris et al. 2004; Johnston and Shearing 2003). The actor skills that are part of the ABM were operationalized as follows.

- **Mentalities** include the extent to which an actor takes a *long-term focused preventive or risk-oriented position* with regard to cocaine trafficking and the extent to which an actor takes a *short-term focused, punishment-oriented position*.
- **Technologies** enable a node to influence security governance. This paper defines these technologies as the ability to *enforce rules or laws* and the ability to *exert pressur*e on other actors in the network.
- **Resources** enable a node to support its operations. In this paper, these resources are *economic*, *cultural*, and *social capital*, as described by Burris et al. (Burris et al. 2004) and derived from Bourdieu (Bourdieu 1986). Based on the empirical research in the PoR, additional resources are included in this paper: the extent to which an actor is driven by *reputation benefits*, the perceived *sense of urgency* to act against cocaine trafficking, the sense of *responsibility to act based on formal rules and regulations*, and the sense of *responsibility to act based on social norms and expectations*.
- **Institutional structure** enables a node to mobilize mentalities, technologies, and resources. For instance, nodes may be *public or private*, have a specific *geographic scope of authority*, and work at an *operational or strategic level* to address cocaine trafficking

To understand how the nodes and their skills relate to each other in the context of nodal governance, a set of overarching relational parameters is used (Sergi et al. 2021). These form the CLD. Building on Sergi et al. (Sergi et al. 2021) and Brewer (Brewer 2014, 2015, 2017) the relational parameters include **trust**, **cooperation** and **competition** between actors. Based on the empirical research in the PoR, we also consider the **enforcement** of rules or laws related to port security by one actor on another and the application of **pressure** to act against cocaine trafficking by one actor on another as relational parameters.

## 1.1 System Architecture and Mechanisms

The structural dependencies within the model are divided into two primary interaction matrices based on domain expertise (see Tables 1 and 2 in the Supplementary Appendix), forming a multi-layer model (see Figure [fig:layers]):

1. **Parameter-to-Parameter Cross-Impacts:** Dictates how a change in one relational parameter directly influences another.

2. **Parameter-to-Skill Cross-Impacts:** Dictates how relational dynamics alter the underlying skills of the actors over time.

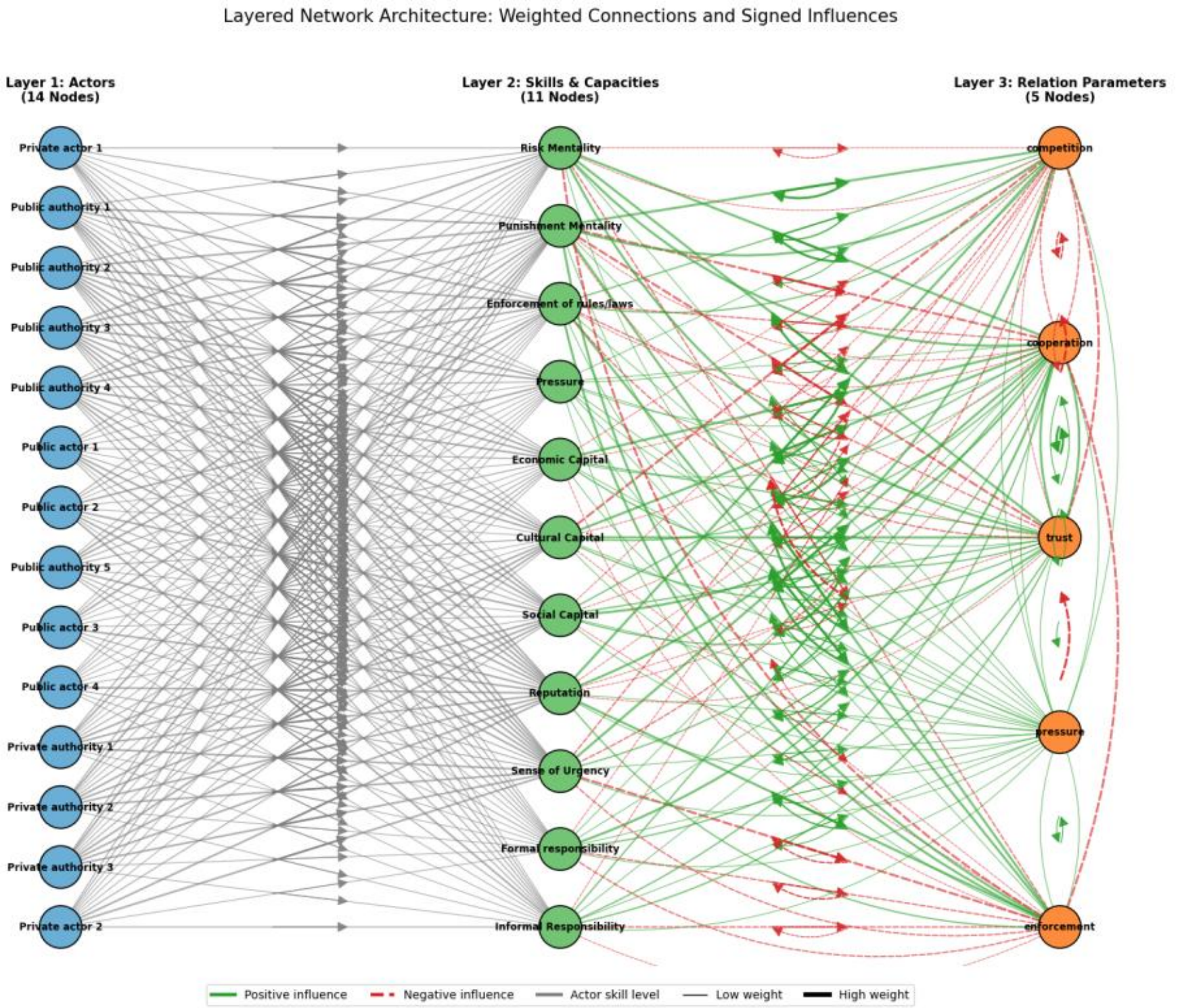

We have chosen to operationalize domain expertise based on the work of Oetker et al. (Oetker et al. 2026), Seidl (Seidl 2014) and Yang and Gilbert (Yang and Gilbert 2008). The network structure is governed by explicit cross-impact coefficients, set by the authors based on empirical insights from the domain expert. Relational parameters dynamically modify actor skills at each simulation step. Shock propagation is therefore intentionally gradual rather than instantaneous. A disturbance to one relational parameter first affects other relational parameters through the parameter-to-parameter cross-impact matrix, then modifies actor skills through the parameter-to-skill interactions, after which changes in skills and relational conditions influence network relationships and aggregate PPC effectiveness. The propagation rates are explicitly parameterized rather than assumed to occur within a single time step: parameter-to-parameter and skill-to-parameter effects use a rate of 0.030, parameter-to-skill effects use a rate of 0.035, while indicator relaxation uses a rate of 0.09 (see Table 5 in the Suplementary Appendix). Consequently, a shock can produce an immediate change in the directly affected parameter while its indirect effects emerge gradually over subsequent simulation steps.

## 1.2 Optimizing Effectiveness and System Rules

The objective of the toy model is to explore how different configurations of actor skills and relational dynamics affect model-defined PPC effectiveness. Here, PPC effectiveness refers to the model-defined capacity of the network to sustain effective cooperation, rather than to a direct empirical measure of reductions in cocaine trafficking or other real-world outcomes. The optimization is governed by a set of behavioral and institutional rules, summarized in the Supplementary Appendix, Table 4. These rules define the mechanisms and constraints that determine whether cooperation is functional. The corresponding numerical parameter values used to operationalize these rules, including propagation rates, pressure thresholds, decay coefficients, and relaxation parameters, are listed in Table 5 in the Supplementary Appendix. The PPC effectiveness score is explained in the Methodology section below. Per the network criteria outlined in Table 4, successful cooperation over time requires satisfying the following rule thresholds:

- **Balanced Dynamics:** Cooperation $>$ Competition, and Trust $> 0$ with a positive growth trajectory ($\Delta$Trust $> 0$).
- **Institutional Composition:** The network must contain at least one public actor, one private actor, one operational-level node, and one actor with direct authority over the PoR scope.
- **Mentalities & Resources:** Nodes must maintain a Risk Mentality that is greater than a Punishment Mentality, and all three capital types (Economic, Cultural, Social) must remain strictly greater than zero.

When external shocks are applied to specific nodes, the changes ripple through the CLD structure. The programmatic implementation embeds these exact constraints, running parametric sweep adjustments to identify the optimum effectiveness of parameter interactions for the model. Table 3 in the Supplementary Appendix logs how the structural rules, constraints, and numerical parameters found within the interactions in the toy model are mapped into the execution of the computational ABM.

# 2 Methodology

The code file of the our model can be found under https://github.com/FOetker/Public-Private-Cooperation-in-the-PoR-An-ABM-of-of-Cooperative-Efforts-Against-Cocaine-Trafficking.git.

## 2.1 Behavioural Rules

Our exploratory model represents PPC as a dynamic socio-institutional system in which relational indicators, actor skills, and institutional characteristics evolve simultaneously through a set of interconnected rules derived from the conceptual framework and accompanying rule base. Each simulation step corresponds to a three-month period during which external shocks, endogenous adaptation (changes generated within the simulated system by the interactions and feedback between actors and model variables), and network interactions determine the systems evolution. Additional mechanisms include homeostatic relaxation towards initial state, gradual decay of enforcement over time, public- and private-sector modifier effects, dynamic network rewiring (connections between actors are not fixed and can change during the simulation, weak relationships can be dissolved when they fall below a specified threshold, and other relationships can be strengthened or newly formed) and externally imposed shocks.

Homeostatic relaxation in our model represents the gradual return of relational parameters and actor skills towards baseline values based on empirical research on the Dutch approach to organised drug crime that shows actors' inability to establish long-term divergence (Nelen et al. 2023). The research also demonstrates how policy priorities can change, resulting in weakened enforcement efforts over time unless further investments are made. In our model, this enforcement decay assumes that enforcement capacity declines proportionally over time unless actively renewed:

$$\begin{equation} \label{eq:enforcement decay} \[ E_{t+1}=E_t(1-\delta_E)+X_{\mathrm{Inputs}}, \qquad \delta_E=0.015, \] \end{equation}$$

so that, without reinforcing inputs, enforcement is multiplied by $0.985$ per quarterly step rather than reduced by a fixed $0.015$. The input term $X_{\mathrm{Inputs}}$ can partially or fully offset this decay.

Public- and private-sector modifiers capture hypothetical differences in actor behavior: public actors can exert greater pressure due to their political leverage, while private actors face greater competition in the market economy (Nelen et al. 2023). Dynamic network rewiring represents hypothetical changes in actor relationships over time, including the dissolution of weak ties below a threshold and network adaptation. External shocks introduce disturbances from outside the model that can affect system conditions and trigger further adaptation. Together, these mechanisms capture gradual endogenous adaptation and structural change within the PPC system. A detailed breakdown is provided in Table 4 in the Supplementary Appendix. Overall system performance is evaluated using a composite PPC effectiveness score, which combines structural validity requirements, pressure optimization, trust and cooperation dynamics, competition–cooperation balance, actor mentalities, available resources, responsibility, and reputation into a normalized performance measure ranging from 0 to 1. This formulation enables the model to capture both the direct effects of parameter changes and the indirect effects arising from feedback loops, network propagation, and adaptive system behavior over time. For a detailed breakdown of the effectiveness calculation in the PPC model, please review equation [eq:ppc_effectiveness] in the Supplementary Appendix.

### 2.1.1 Relational dynamics

The relational core is operationalized as a signed CLD (Figure 3 in the Supplementary Appendix), with five relational parameters as nodes and signed directed edges indicating hypothesized causal direction and relative strength. Nodes represent variables or relational parameters; edge signs indicate whether an increase in one variable increases or decreases another, while weights represent assumed influence strength. The primary reinforcing loop (R) links trust and cooperation, while balancing loops (B) through competition and enforcement constrain trust and organic tie formation. Relational parameters incorporate homeostatic relaxation and multiplicative decay to represent institutional friction and operational erosion. Enforcement decays multiplicatively by 1.5% per quarterly simulation step (factor 0.985) without continuous reinforcement or external inputs, reflecting the hypothetical weakening of enforcement capacity and deterrence; this rate is a modeling assumption rather than an empirical estimate. Trust and cooperation reinforce each other and, following domain-expert hypotheses, require positive levels to increase, whereas competition limits cooperation, which is effective only when cooperation exceeds competition. Further details on enforcement and pressure dynamics are provided in the Supplementary Appendix; CLD development follows Crielaard et al. (Crielaard et al. 2024) and Burns and Musa (Burns and Musa 2001).

### 2.1.2 Actor-specific institutional effects

Institutional characteristics determine actor influence. Effective cooperation hypothetically requires public and private actors, an operational actor, and port-level authority. Public actors receive stronger enforcement and pressure effects, while private actors exhibit stronger competition effects (Nelen et al. 2023); these magnitudes are modeling assumptions given limited empirical backing. Cooperation also requires a risk-oriented mentality, sufficient economic, cultural, and social capital, and urgency, while actor attractiveness depends on these skills and formal/informal responsibility. Due to limited expert specifications, equal weighting and update rules are assumed. Table 4 in the Supplementary Appendix summarizes the implemented rules, interpretations, and origins.

### 2.1.3 PPC Effectiveness

PPC effectiveness is a model-defined score ($0$–$1$) representing the system's capacity to sustain effective public-private cooperation. It combines structural, relational, actor, resource, responsibility, and reputation factors and is recalculated each simulation step within $[0,1]$. It is a model output, not a direct measure of cocaine-trafficking reduction or other real-world outcomes; its weighted aggregation uses endogenous outputs of nonlinear, interacting mechanisms (see PPC Effectiveness equation [eq:ppc_effectiveness]). Robustness is assessed through exhaustive shock experiments across all relational-parameter combinations. Shocks range from $-1.0$ to $1.0$, with resulting values constrained to $[0,1]$. Each configuration is simulated for 5 years and compared with the corresponding unshocked baseline.

## 2.2 Shocking of parameters

The OAT shock analysis tests isolated parameter effects across three baselines: nullpoint (all relational parameters = 0), balance (all = 0.5), and heterogeneous, reflecting expert insights on early PPC development (competition 0.8, enforcement 0.7, pressure 0.5, cooperation 0.4, trust 0.3). Each parameter receives shocks of $\pm 0.15$ and $\pm 0.30$, with outcomes simulated over five years against baseline PPC effectiveness. The exhaustive combinatorial analysis then optimizes interventions from the heterogeneous baseline, testing all relational-indicator combinations with independent shocks from $\pm 0.1$ to $\pm 1.0$ in 0.1 increments, constrained to $[0,1]$. It simulates 161,050 unique scenarios over five years, comparing final PPC effectiveness with the unshocked baseline. Shock magnitudes represent changes applied to the initial parameter values rather than absolute parameter levels. For example, a +0.7 trust shock applied to the heterogeneous baseline increases trust from 0.3 to the upper bound of 1.0. The extensive shock experiments provide robustness and exploratory analysis rather than empirical validation, testing whether qualitative conclusions depend on selected scenarios without demonstrating reproduction of observed real-world outcomes. The ABM is therefore used primarily for mechanism exploration and scenario analysis, not prediction of enforcement statistics or directly measured trafficking reductions. Empirical validation against observed PPC trajectories and additional cases remains an important next step in model development.

# 3 Results

## 3.1 Shocks

### 3.1.1 One-At-a-Time Analysis

We test optimal modeled PPC effectiveness by applying OAT shocks to the toy model under different initial conditions. In the low-capacity null-point and balanced configurations, positive and negative shocks rapidly dissipate toward baseline levels, indicating that isolated disturbances cannot permanently shift low-capacity persistent states. In contrast, the heterogeneous configuration shows sustained network-wide effects: positive shocks to high-leverage drivers, particularly cooperation and trust, maintain elevated effectiveness, while negative shocks cause immediate declines. Across all scenarios, OAT shocks reveal distinct temporal dynamics and parameter leverage. As shown in Figure 5, 6 and 7 in the Supplementary Appendix null-point and balanced configurations progressively increase in effectiveness, reaching approximately 0.30–0.50 by Year 5, showing that the system remains dynamically responsive to targeted perturbations.

Individual parameter shocks ($\pm 0.30$) identify cooperation as the most effective single intervention (Figure 1), increasing PPC effectiveness to 0.456 ($\Delta = +0.134$ from baseline). Negative shocks consistently reduce performance across indicators, with declines in relational parameters producing rapid and persistent losses. Over the five-year OAT simulations, institutional drivers show the strongest baseline leverage, with enforcement ($r = 0.429$) and pressure ($r = 0.335$) having the highest positive correlations with PPC effectiveness (Figure 13 in the Supplementary Appendix). However, subsequent combinatorial analysis shows that enforcement's positive effect depends on independent adjustment; within multi-parameter configurations, selective negative shocks become important for reducing system friction.

### 3.1.2 Multi-parameter Analysis

The multi-parameter analysis identifies several intervention strategies that substantially outperform both the heterogeneous baseline (PPC effectiveness = 0.323) and the simpler uniform-shock experiments. Notably, all of the highest-performing scenarios involve simultaneous intervention across all five relational indicators. The optimal intervention combined a reduction in competition ($-0.6$) and enforcement ($-0.5$) with increases in cooperation ($+0.6$), trust ($+0.7$), and pressure ($+0.5$). This intervention increases PPC effectiveness in the model from 0.323 to 0.568 (+0.246). Similar intervention profiles occupy the remaining top-ranking positions, differing only slightly in the exact shock magnitudes while consistently favoring lower competition and enforcement alongside higher cooperation, trust, and pressure (see Figure 25 in the Supplementary Appendix).

The best-performing intervention is identified for each possible number of simultaneously shocked network parameters (for a comparison to the heterogeneous baseline, see Figure 12 in the Suppelementary Appendix). Figure 1 presents the optimum PPC effectiveness achieved within each intervention size. An increase in attainable effectiveness is observed as additional parameters are incorporated into the intervention strategy. The best single-parameter intervention consisted of increasing cooperation by 0.3, yielding a PPC effectiveness of 0.456 (an improvement of 0.134 over the heterogeneous baseline of 0.323). This denotes a difference of 0.11 to the best overall intervention, which simultaneously adjusted all five network parameters with a resulting PPC effectiveness of 0.568.

Lastly, a sensitivity analysis across all shock combinations in the heterogeneous baseline reveals clear hierarchies in parameter influence (Figure 13 in the Supplementary Appendix). Institutional mechanisms exert the strongest positive impact on effectiveness, led by enforcement ($r = 0.429, \rho = 0.516$) and pressure ($r = 0.335, \rho = 0.406$). Conversely, relational drivers show substantially weaker linear and monotonic correlations, with cooperation ($r = 0.142, \rho = 0.158$), trust ($r = 0.136, \rho = 0.150$), and competition ($r = 0.118, \rho = 0.150$) exhibiting smaller overall gains across the scenario space.

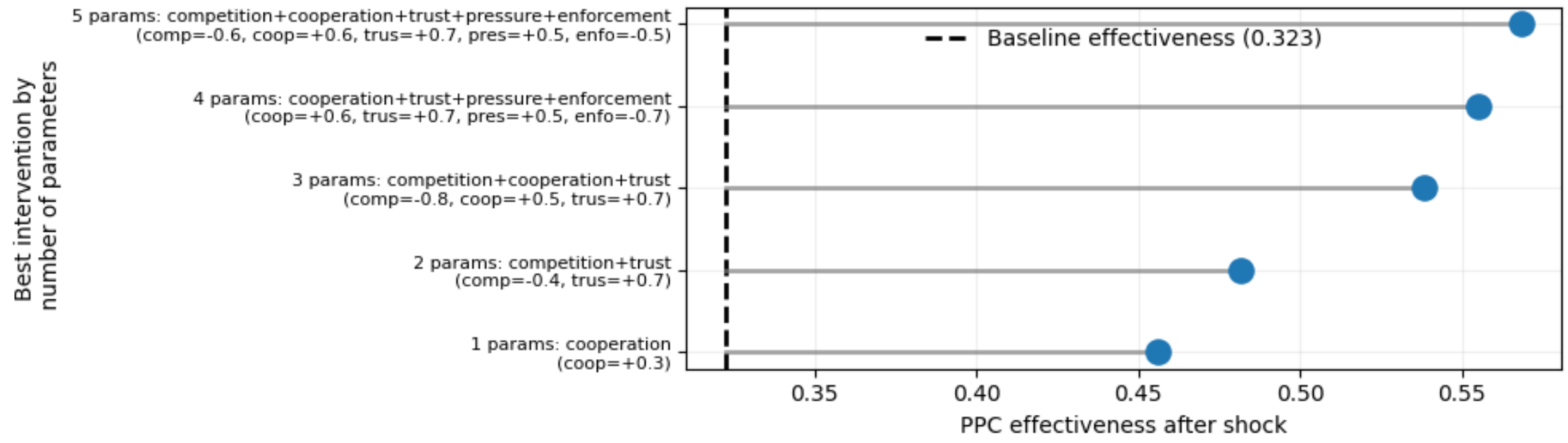

*Optimum PPC effectiveness for each intervention complexity (one to five simultaneously shocked network parameters) under the heterogeneous initial scenario. Each point represents the highest-performing intervention for that size. Dashed line indicates baseline PPC effectiveness.*

### 3.1.3 Enforcement dynamics

We highlight how key model rules shape system outcomes. Across shock scenarios, enforcement follows a non-renewal decay rule, with a multiplicative factor of 0.985 per quarterly simulation step. Thus, in the absence of compensating inputs, enforcement declines proportionally by 1.5% of its current value at each step rather than by a fixed absolute amount. This cumulative decay contributes to the downward trajectories observed in Figures [fig:heterogeneous_shocks], 19, and 24. Endogenous propagation mechanisms do not fully offset this decay, limiting the long-term impact of initial shock types.

This trajectory reflects the model's influence structure. Enforcement receives positive input only from pressure (weight = 0.5) and reinforcement from enforcement-related skills, punishment mentality, and reputation, whereas risk mentality, sense of urgency, and formal/informal responsibility exert negative effects. Because these feedback loops provide limited compensation for decay, this pattern indicates not an insensitivity to shocks, but a operationalization that prioritizes enforcement erosion over adaptive renewal. The initially sharp decline of the heterogeneous baseline is therefore not an emergent empirical finding that enforcement naturally declines at this rate, but a consequence of the assumed non-renewal mechanism. Because enforcement is multiplicatively decayed at each quarterly step and endogenous reinforcement is initially insufficient to compensate for this decay, the baseline trajectory declines rapidly during the early simulation period.

# 4 Discussion

### 4.0.1 One-At-a-Time Analysis

The OAT shock analysis reveals two hypothetical governance implications: state-dependent inertia and asymmetric parameter leverage. Isolated relational interventions depend on initial structural capacity: in the low-capacity null-point and balanced configurations, single-parameter shocks produced no long-term improvement in PPC effectiveness ($\Delta = 0.00$). This leaves these systems trapped in low-performance equilibria and indicating that multiple relational dimensions must be targeted simultaneously to overcome structural inertia. In the heterogeneous configuration, relational parameters had unequal effects: positive shocks to cooperation and trust increased PPC effectiveness by up to $\Delta = +0.134$, while disturbances to competition and enforcement produced little or no lasting change. Consistent with previous findings from for example Brewer(Brewer 2014) and Quéro and Dupont(Quéro and Dupont 2019), cooperation and trust act as high-leverage variables whose network effects sustain system-wide improvements. Other parameters have limited influence. According to the model governance efforts should consequently prioritize building trust and reducing competitive friction simultaneously, as isolated adjustments are insufficient to overcome structural inertia. This hypothesis should be empirically validated.

### 4.0.2 Multi-parameter Analysis

Combinatorial multi-parameter optimization reveals non-linear synergies that outperform isolated interventions (Table 6, Figure 25 in the Supplementary Appendix). The top strategy of simultaneous shocks across five parameters (-0.6, 0.6, 0.7, 0.5, and -0.5) achieves peak PPC effectiveness of 0.568, improving 0.2459 over the unshocked baseline (0.322) and 0.1123 over the best single-parameter strategy (Table 6 in the Supplementary Appendix). Over the simulation span, the heterogeneous baseline is outperformed by the best performing shock (see Figure 12). Combined interventions are therefore more effective up to a point, with diminishing returns beyond three or four simultaneous adjustments, making further increases in intervention number or magnitude potentially unjustified given their resource requirements according to the model.

Notably, optimal configurations prescribe substantial reductions in enforcement ($-0.4$ to $-0.5$) alongside aggressive competition cuts ($-0.6$ to $-0.8$). Although bivariate sensitivity analysis links higher enforcement to effectiveness under static conditions, multi-parameter optimization shows that over-regulation creates friction that inhibits cooperation. Hypothetically, therefore, sustainable PPC efforts require shifting from top-down enforcement efforts of rules and regulations toward trust-building, as rigid compliance mechanisms increase friction among actors. Maximizing PPC performance consequently would hypothetically require lower-friction, high-cooperation environments rather than high-enforcement, high-competition structures. Previous research on port security points to a similar conclusion (Brewer 2014). The optimization confirms that multi-parameter coordination is essential: single-lever adjustments are absorbed by institutional friction, whereas simultaneous changes across relational and institutional boundaries can break systemic inertia.

### 4.0.3 Sensitivity of Parameters

Considering the sensitivity analysis as displayed in Figure 13 in the Supplementary Appendix, PPC effectiveness in the model is initially driven primarily by formal institutional mechanisms, while relational dynamics gain importance over time. However, these correlations capture isolated parameter effects and do not reflect interactions between parameters. The combinatorial optimization analysis therefore reveals that coordinated multi-parameter changes can produce different outcomes due to non-linear system dynamics. Thus, the effectiveness index should be interpreted as a summary measure of outcomes emerging from the model's non-linear and dynamic processes, rather than as a standalone linear representation of the underlying ABM relationships.

# 5 Conclusion

This study develops an Agent-Based toy Model (ABM) to explore how relational dynamics hypothetically influence Public-Private Cooperation (PPC) against cocaine trafficking in the Port of Rotterdam (PoR). Consistent with previous research on PPC and port security (Brewer 2014; Quéro and Dupont 2019; Waardenburg et al. 2016), we found that the highest level of modeled effectiveness is achieved through multi-parameter interventions that simultaneously strengthened cooperative mechanisms while reducing competitive and enforcement-related friction. This hypothetically implies that coordinated, multi-lever interventions would therefore outperform isolated policy adjustments.

Several future research directions remain. First, the empirical foundation should be strengthened by involving additional domain experts through workshops, Delphi studies, or participatory model building to refine assumptions about effect sizes, update speeds, and decay mechanisms (see Crielaard et al. (Crielaard et al. 2024), Franco and Montibeller (Franco and Montibeller 2010), and Norström et al. (Norström et al. 2020)). Second, future work should move beyond exploratory shock experiments towards realistic, actor-specific governance interventions that account for heterogeneity in decision-making, organizational objectives, resource constraints, and learning. Third, validation against additional empirical PPC cases and longitudinal data, alongside renewal mechanisms such as investment, learning, policy change, and repeated cooperation, would improve the representation of how governance capacities evolve over time. The Supplementary Appendix provides proposed pattern-oriented model checks (Table 7) to support comparison of model-generated outcomes with real-world patterns. This addresses the final stage of model building described by Oetker et al. (Oetker et al. 2026), which would move the model from calibration towards validation.

# 6 Acknowledgements

This research is part of the project Focusing on the Right Things in the Port of Rotterdam (with project number KICH1.VE03.21.008) of the research programme KIC-MISSION 2021 which is (partly) financed by the Dutch Research Council (NWO).

# Appendix

The Supplementary Appendix is available at: https://github.com/FOetker/Public-Private-Cooperation-in-the-PoR-An-ABM-of-of-Cooperative-Efforts-Against-Cocaine-Trafficking.git.

# 7 Supplementary Appendix

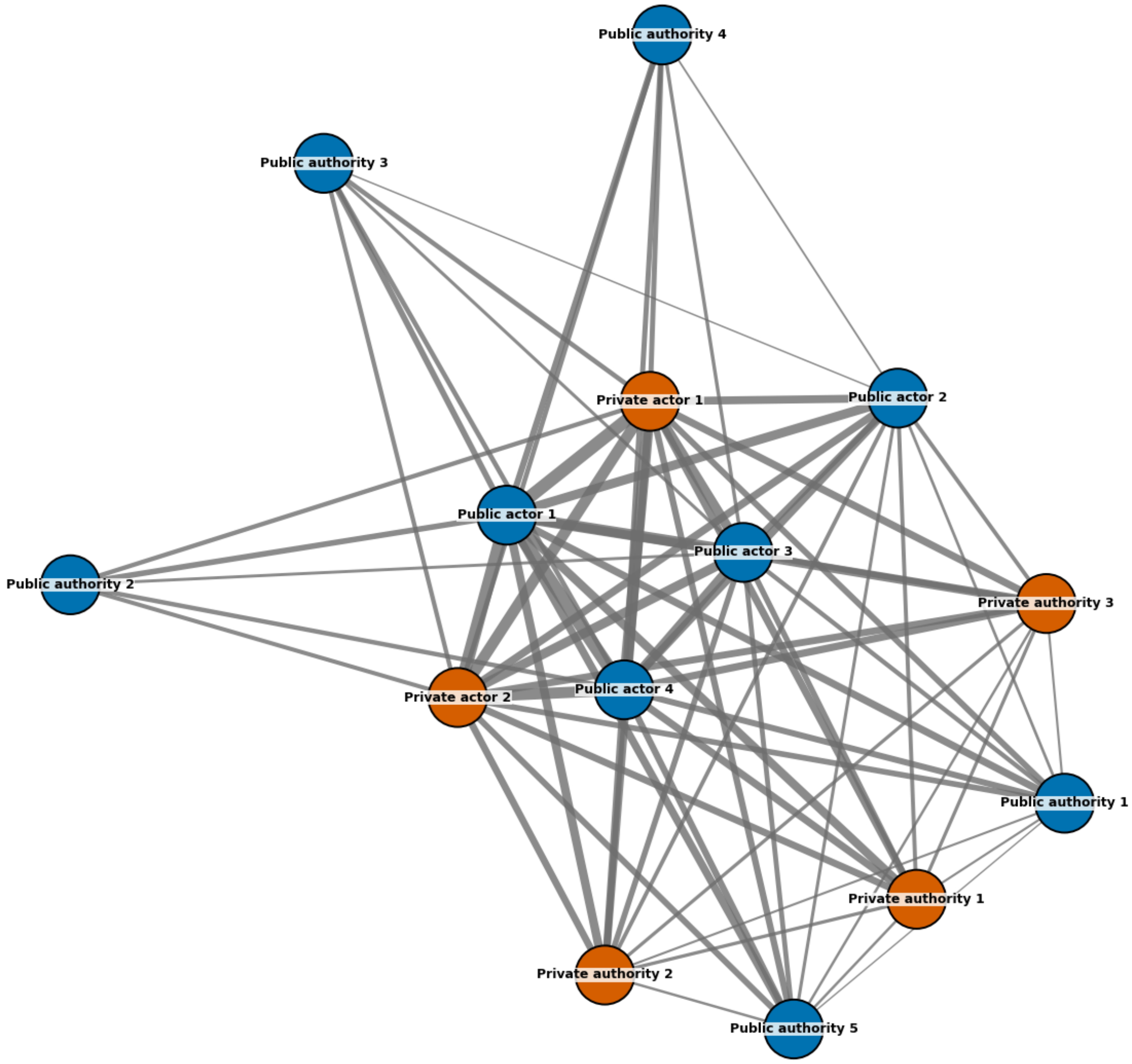

*Baseline actor network generated through endogenous tie formation. Nodes represent actors, with blue nodes indicating public actors and orange nodes indicating private actors. Edges represent simulated collaborative ties derived from actor skill profiles, where thicker edges indicate stronger tie strength. The network contains 14 actors and 72 ties.The network density is 0.79, with an average clustering of 0.8 and an average degree of 10.3. The average tie strength is 0.16, and the most central actor is Public Actor 1.*

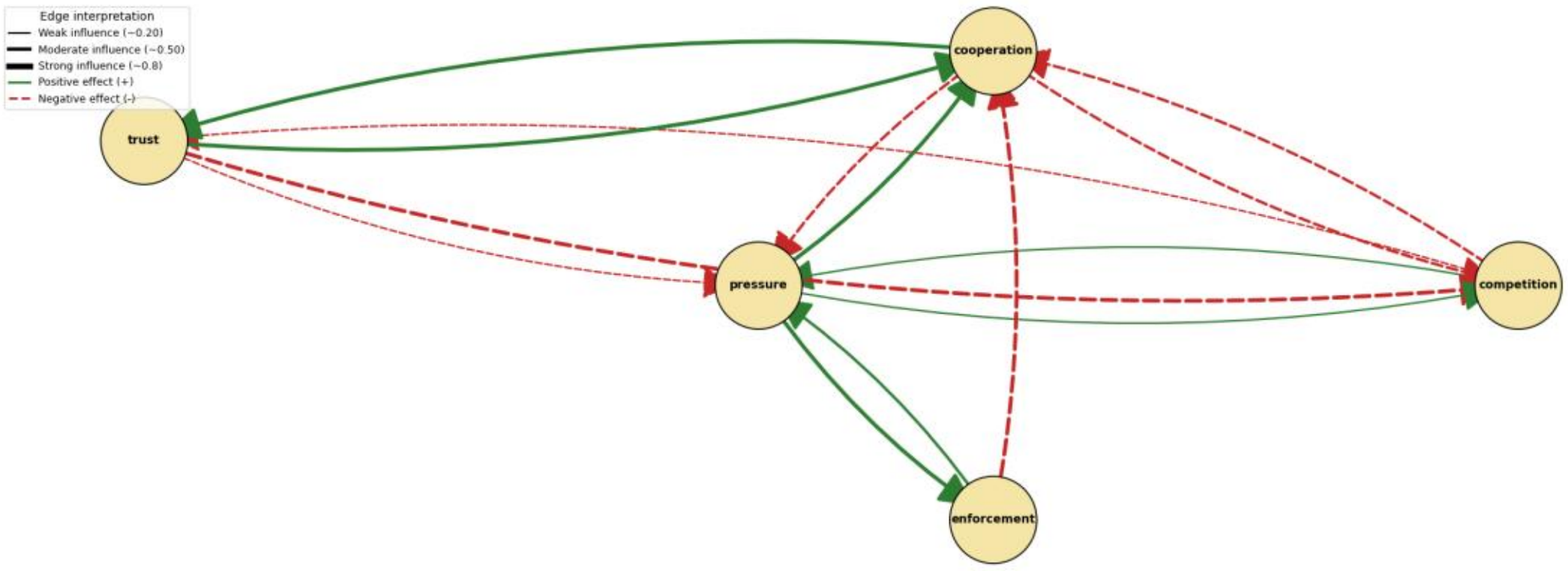

*CLD of the relation-parameter core of the PPC model. Nodes represent relational parameters, with green solid arrows indicating positive effects and red dashed arrows negative effects.*

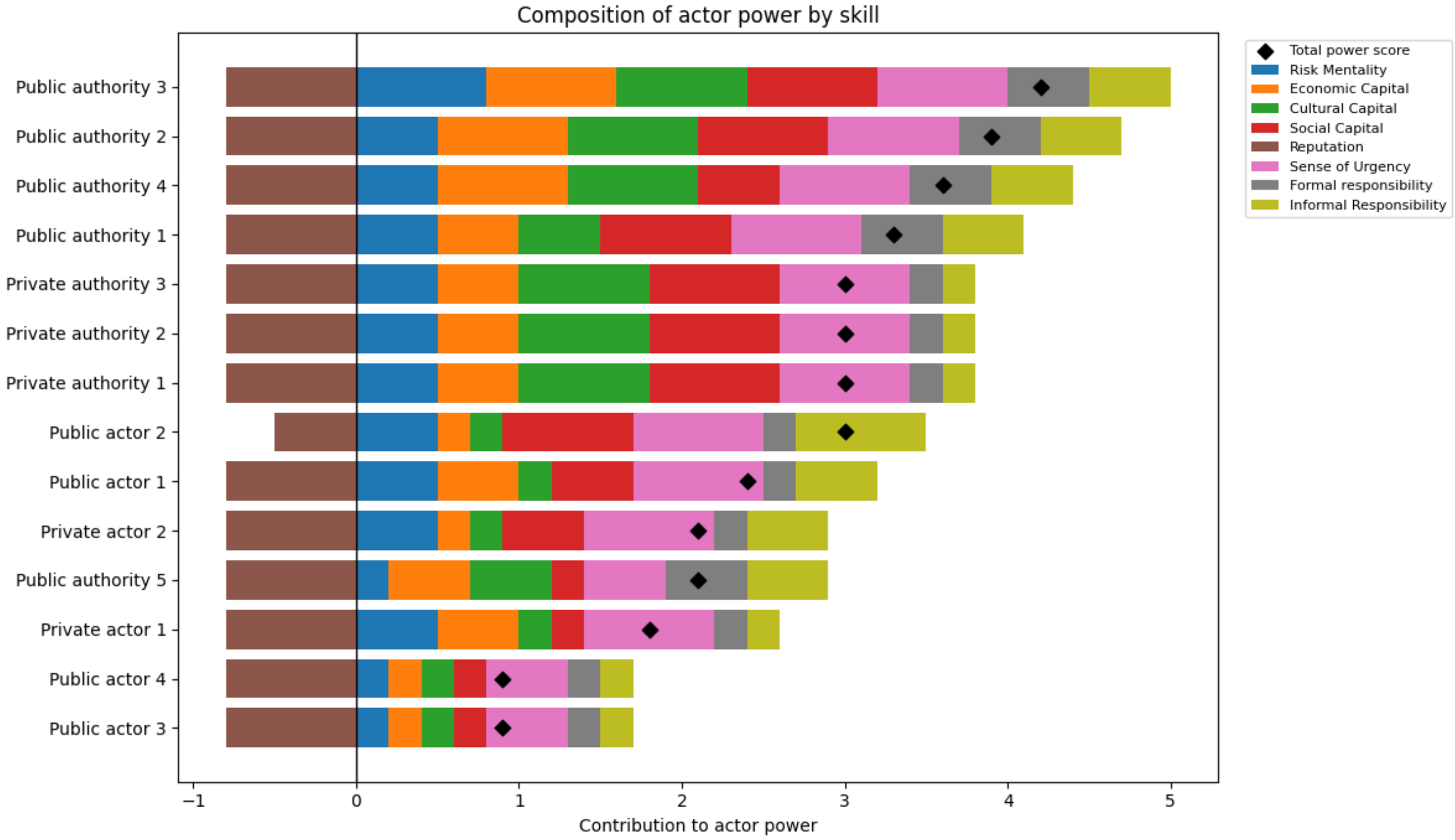

*Actor skills and institutional characteristics used in the ABM initialization. Values represent normalized skill levels between 0 and 1. Skills may change over time through model dynamics. Reputation accounts negatively for the cummulative actor skills. The skills were categorized in low, medium and high.*

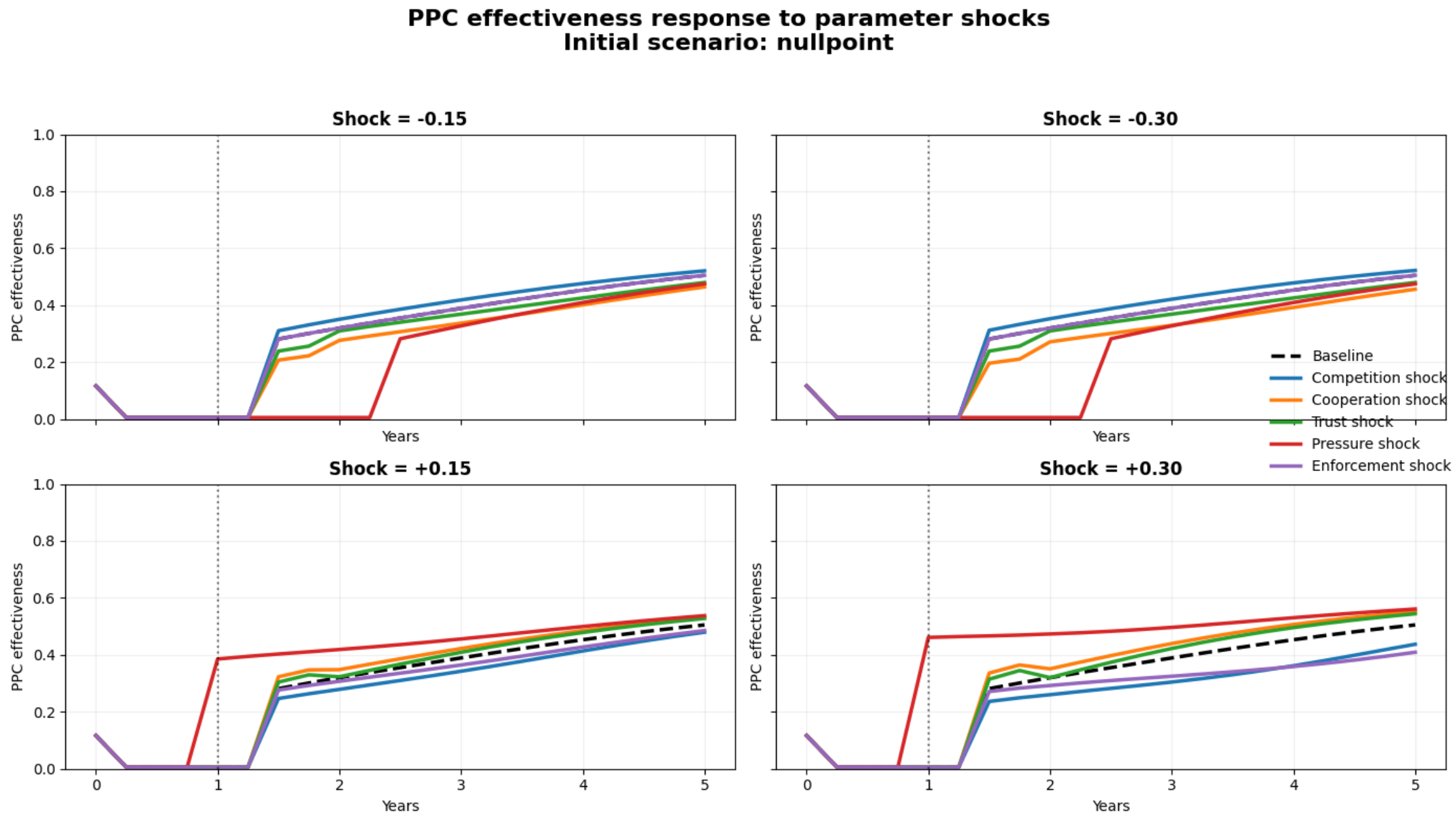

*PPC effectiveness following single-parameter shock interventions under the nullpoint initialization scenario. All relational parameters are initialized at zero, representing the absence of pre-existing relationships. Each line represents the PPC effectiveness trajectory over the five-year simulation period following a single positive or negative parameter shock.*

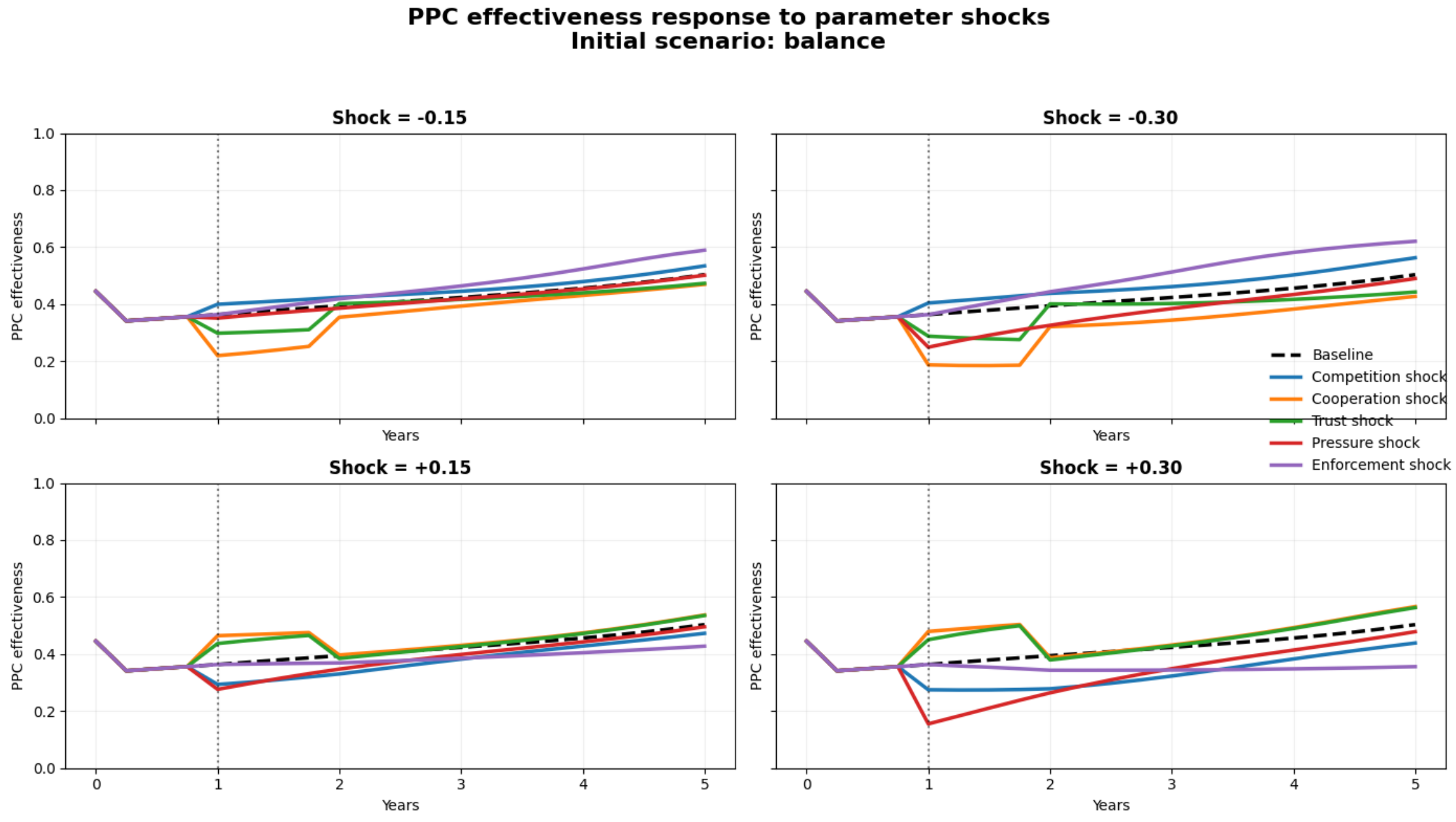

*PPC effectiveness following single-parameter shock interventions under the balanced initialization scenario. All relational parameters are initialized at a neutral value of 0.5, allowing the isolated propagation of shocks through the network from an initial state. Each line represents the PPC effectiveness trajectory over the five-year simulation period following a single positive or negative parameter shock.*

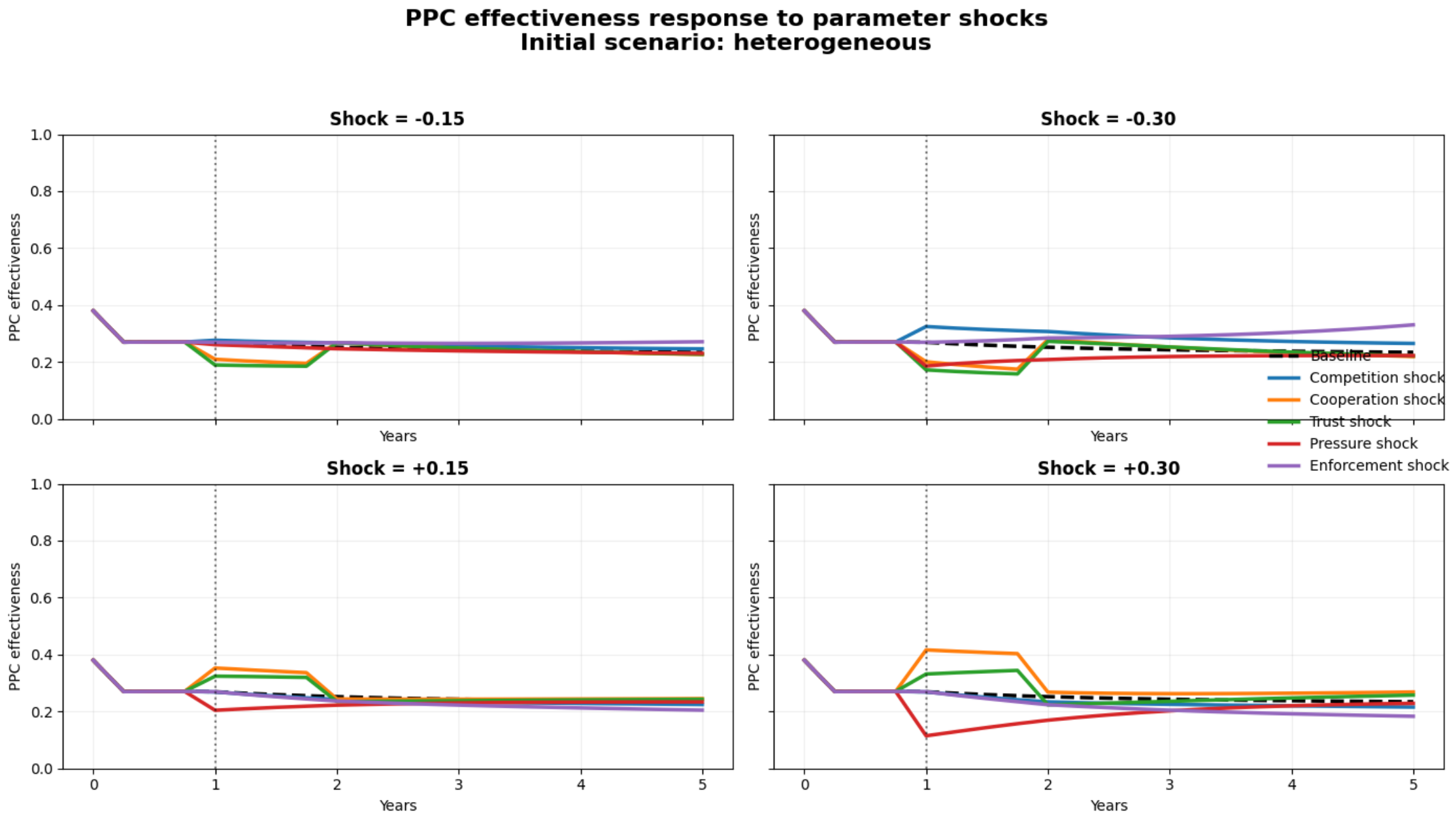

*PPC effectiveness following single-parameter shock interventions under the heterogeneous initialization scenario. Parameter values are derived from expert knowledge, representing a more realistic network configuration with existing relational differences. Each line represents the PPC effectiveness trajectory over the five-year simulation period following a single positive or negative parameter shock.*

PPC effectiveness is not a directly observed variable but a composite performance index that summarizes whether the conditions required for effective Port Public–Private Cooperation (PPC) are satisfied. It is recalculated at every simulation step and constrained to the interval $[0,1]$, where 0 represents the lowest attainable modeled effectiveness and 1 represents the highest attainable modeled effectiveness under the implemented rules.

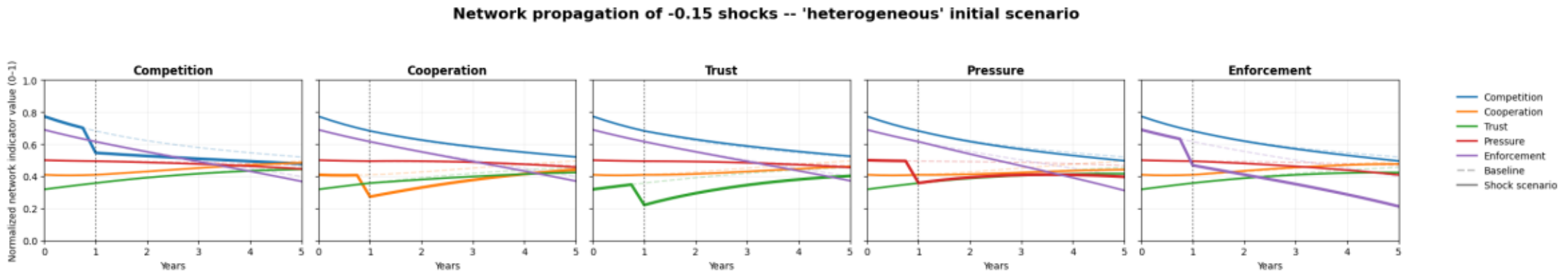

$-0.15$ *shock*

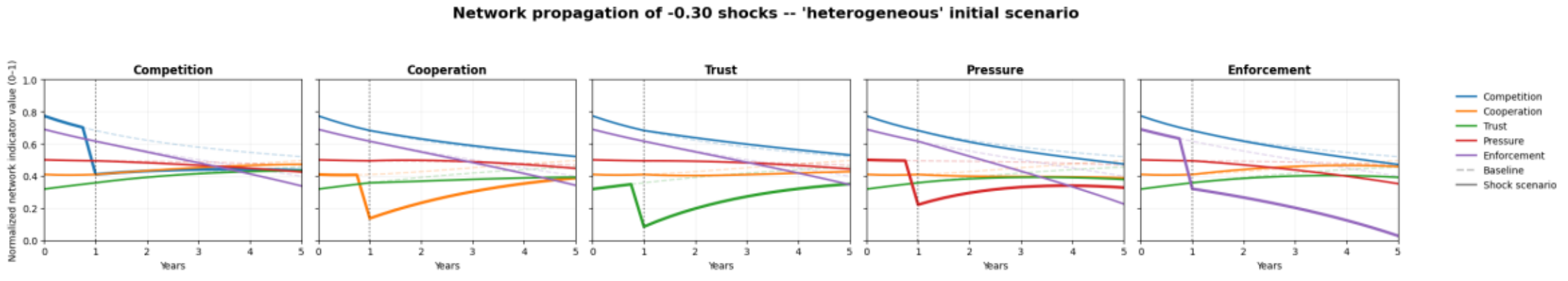

$-0.30$ *shock*

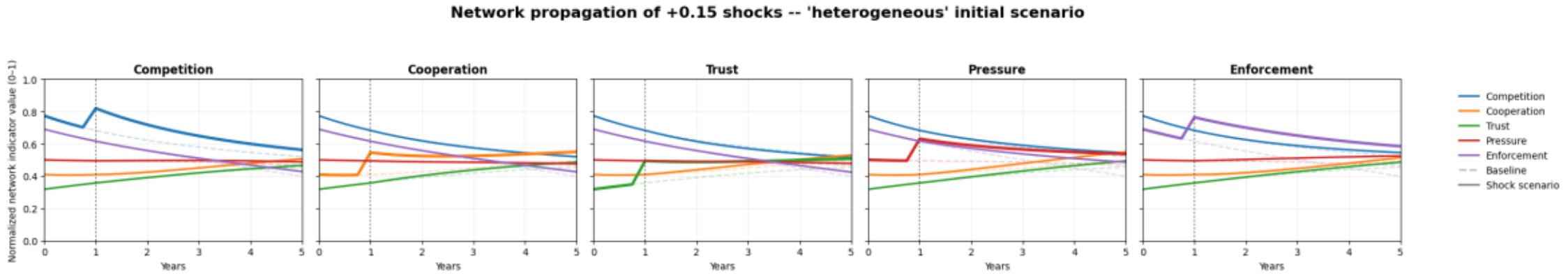

+0.15 *shock*

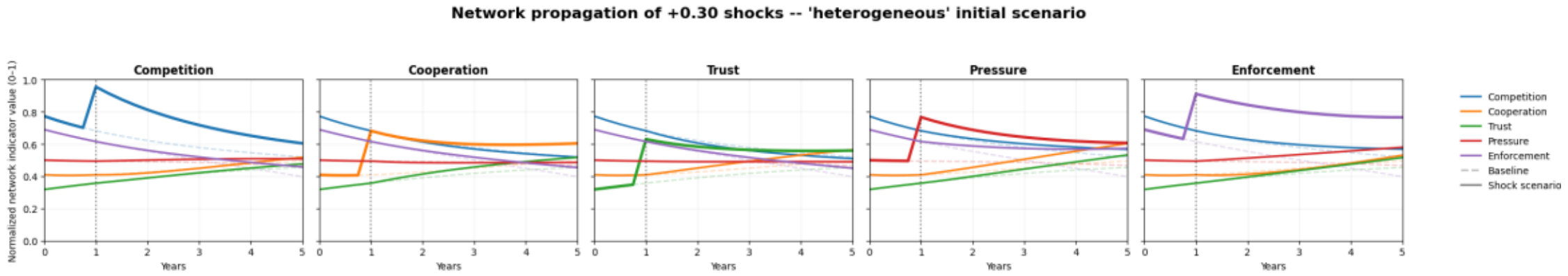

+0.30 *shock*

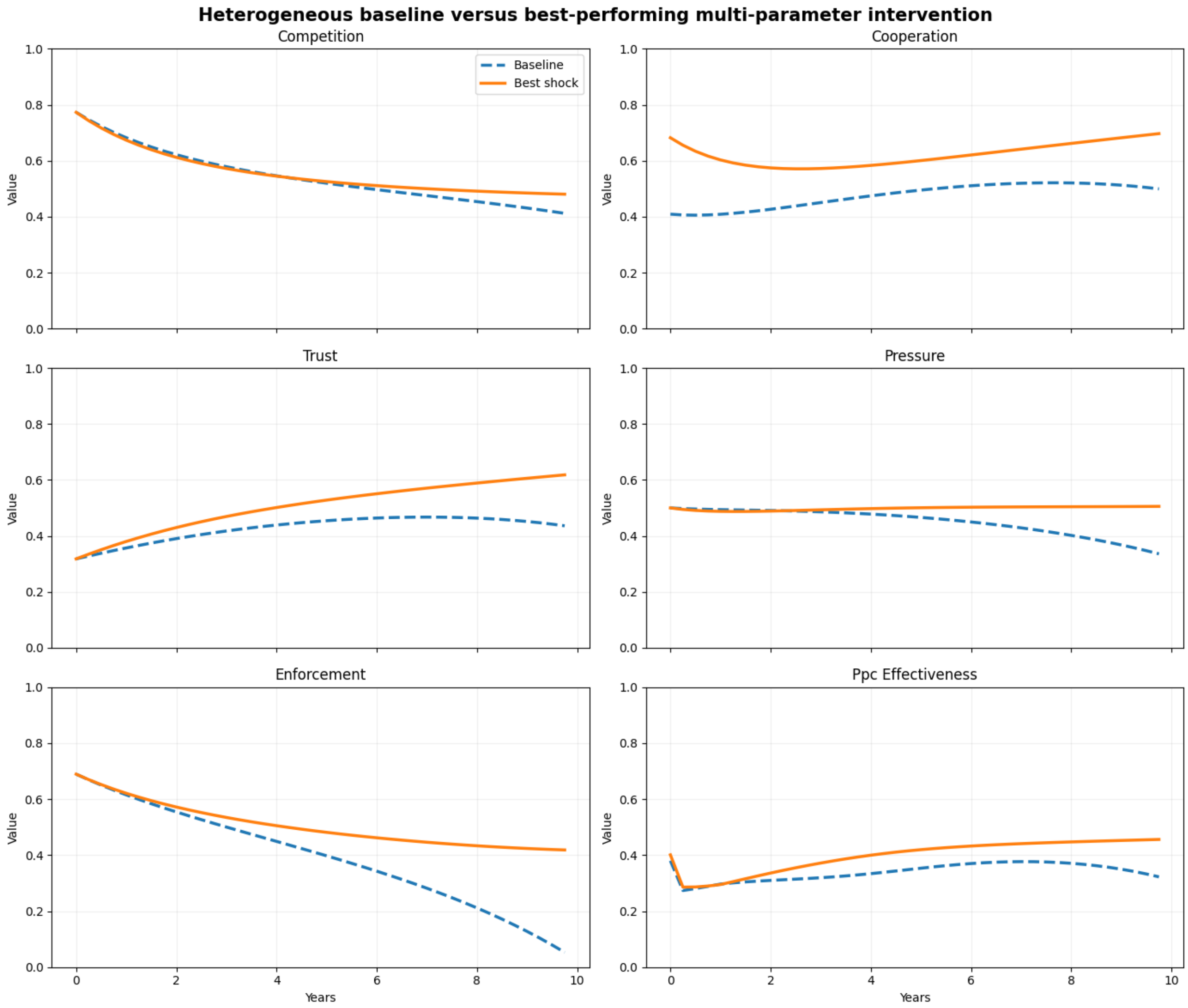

*For the 5 parameters and PPC effectiveness, the heterogeneous baseline scenario is displayed against the best performing multi-parameter intervention. The baseline is shown in blue, with the best performing multi parameter shock in orange.*

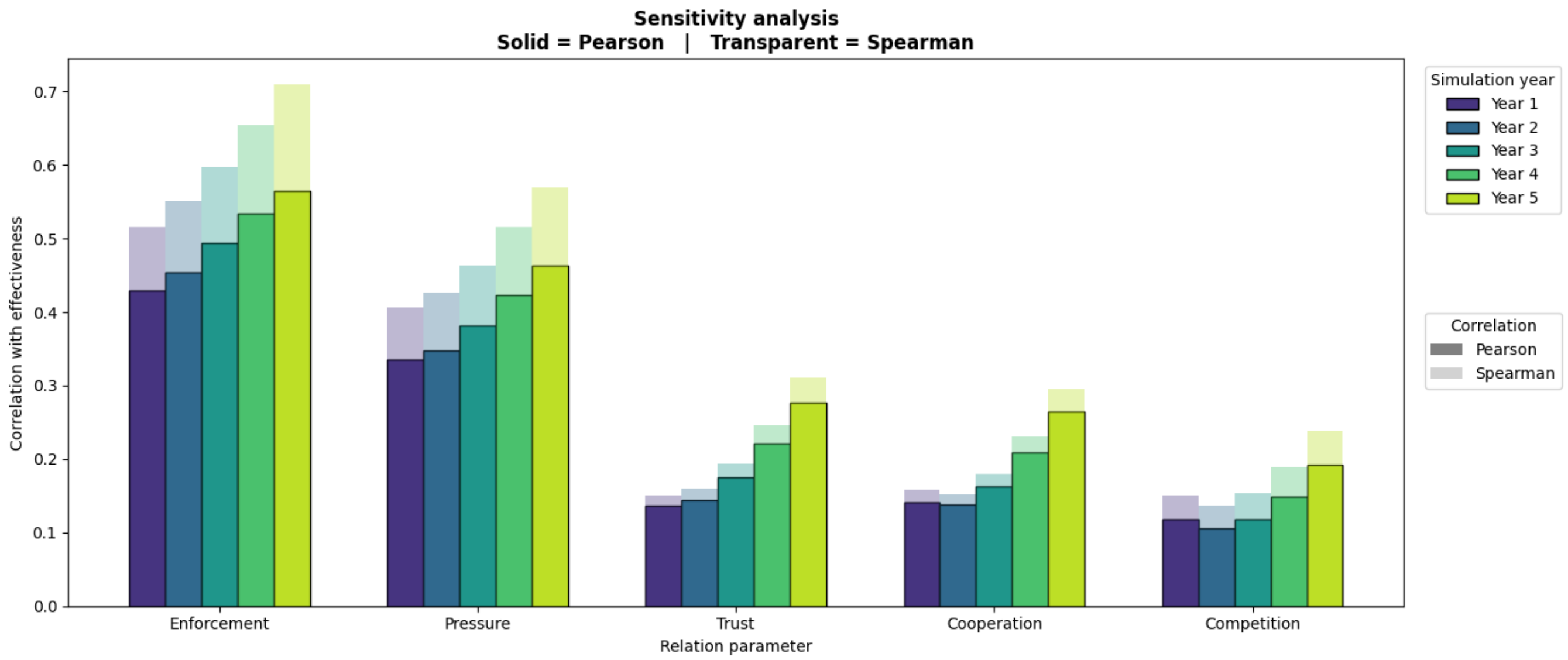

*Parameter sensitivity across shock combinations in the heterogeneous baseline. Institutional mechanisms dominate effectiveness outcomes, with enforcement and pressure showing the strongest associations ($r = 0.429$, $\rho = 0.516$; $r = 0.335$, $\rho = 0.406$), while relational mechanisms exhibit weaker overall influence, as reflected in smaller correlations for cooperation, trust, and competition.*

*Relational Parameter Cross-Impact Matrix*

| Influencing Parameter | competition | cooperation | trust | pressure | enforcement |
|---|---|---|---|---|---|
| **competition** | x | 0.5 | 0.2 | 0.2 | x |
| **cooperation** | 0.5 | x | 0.8 | 0.5 | x |
| **trust** | 0.8 | 0.8 | x | 0.2 | x |
| **pressure** | 0.2 | 0.8 | x | x | 0.8 |
| **enforcement** | x | 0.8 | x | 0.5 | x |

*Relational parameter-to-actor skill influence matrix. Skills include Risk Risk Mentality, Punishment Mentality, Enforcement of rules/laws, Pressure, Economic Capital, Cultural Capital, Social Capital, Reputation, Sense of Urgency, Formal responsibility, Informal Responsibility. Skills include Competition, Cooperation, Trust, Pressure, Enforcement.*

| Parameter | Risk | Punish. | Enforce. | Press. | Econ. | Cult. | Social | Reput. | Urg. | Formal | Informal |
|---|---|---|---|---|---|---|---|---|---|---|---|
| Competition | 0.2 | 0.8 | 0.5 | 0.8 | 0.2 | 0.2 | 0.2 | 0.8 | 0.5 | x | x |

| Parameter | Risk | Punish. | Enforce. | Press. | Econ. | Cult. | Social | Reput. | Urg. | Formal | Informal |
|---|---|---|---|---|---|---|---|---|---|---|---|
| Cooperation | 0.8 | 0.2 | 0.2 | 0.5 | 0.2 | 0.5 | 0.8 | 0.2 | 0.5 | 0.8 | 0.5 |
| Trust | 0.8 | 0.2 | 0.5 | 0.5 | 0.5 | 0.5 | 0.5 | 0.2 | x | 0.5 | 0.5 |
| Pressure | 0.5 | 0.8 | 0.2 | 0.8 | 0.8 | 0.5 | 0.8 | 0.8 | 0.8 | 0.8 | 0.8 |
| Enforcement | 0.8 | 0.8 | 0.8 | 0.5 | 0.2 | x | x | 0.5 | 0.5 | 0.5 | 0.2 |

*ABM and CLD Implementation Traceability Log*

| Rule / System Component | Type | Influenced Elements | Mapping Logic / Assumptions |
|---|---|---|---|
| Parameter Matrix | relation parameters | Cross-parameter loops | Direct coefficient transfer into `param_to_param` structures. |
| Skill Influence Matrix | relation parameters | Actor skill values | Bounded matrix operations constrained to $[0,1]$. |
| Pressure Sweet Spot | rules | Global network cohesion | Evaluation of total rule compliance through parametric shock sweeps. |
| Time Discretization | rules | Model step updates | Verbatim definition where $1 \text{ step} = 3 \text{ months}$ (Quarterly increments). |
| Mentality Disparity | rules | Risk vs. Punishment | Condition checked via a boolean evaluator inside the success loop. |
| Capital Sufficiency | rules | Asset availability | Assumed numeric floor ($> 0.1$) required to satisfy network health. |

*Overview of implemented ABM rules and parameter origins.*

| Rule / Parameter | Implementation | Origin |
|---|---|---|
| Skill-to-parameter influence | Actor skills gradually influence relational parameters (e.g., cooperation and trust). | DE (direction), assumption (rate) |
| Parameter-to-skill influence | Relational parameters influence actor skills over time. | DE (direction), assumption (rate) |
| Parameter feedback | Trust, cooperation, competition, pressure, and enforcement influence one another. | DE (direction), assumption (speed) |
| Minimum pressure requirement | Effective cooperation requires pressure $>$ $0.2$. | DE |

| Rule / Parameter | Implementation | Origin |
|---|---|---|
| Pressure optimum | Cooperation is highest at moderate pressure levels. | DE, assumption (magnitude) |
| Public pressure advantage | Public actors apply pressure more effectively. | DE, assumption (magnitude) |
| Private competition advantage | Private actors exhibit stronger competitive behaviour. | DE, assumption (magnitude) |
| Institutional requirements | At least one public, private, authoritative, and operational actor is required. | DE |
| Risk mentality requirement | Risk mentality must exceed punishment mentality. | DE |
| Resource requirements | Economic, cultural, and social capital, and urgency are required. | DE |
| Relationship dissolution | Very weak network ties are removed below a threshold. | Assumption |
| Relationship relaxation | Relational parameters gradually return to initial state. | Assumption |
| Skill relaxation | Actor skills gradually return to baseline values. | Assumption |
| Enforcement decay | Enforcement decreases each simulation step unless renewed. | DE, assumption (rate) |
| Reputation penalty | Higher average reputation reduces PPC effectiveness. | DE, assumption (weight) |

*Simulation parameter values used in the ABM.*

| Parameter | Value | Description |
|---|---|---|
| Skill → parameter rate | 0.030 | Skill-to-parameter propagation speed |
| Parameter → skill rate | 0.035 | Parameter-to-skill propagation speed |
| Parameter feedback rate | 0.030 | Parameter interaction speed |
| Enforcement decay multiplier | 0.015 | Multiplicative decay per simulation step |
| Reputation penalty weight | 0.15 | Penalty applied to effectiveness |
| Public enforcement boost | 1.15 | Public actor enforcement multiplier |

| Parameter | Value | Description |
|---|---|---|
| Public pressure boost | 1.15 | Public actor pressure multiplier |
| Private competition boost | 1.15 | Private actor competition multiplier |
| Pressure optimum | 0.50 | Centre of inverted-U response |
| Pressure width | 0.28 | Width of effective pressure region |
| Pressure floor | 0.20 | Minimum pressure threshold |
| Tie dissolution threshold | 0.15 | Minimum network tie strength |
| Indicator relaxation | 0.09 | Mean reversion of relation indicators |
| Skill relaxation | 0.05 | Mean reversion of actor skills |

**Enforcement decay:**

$$E_{t+1} = E_t \times (1 - \delta_E) + \sum \text{Inputs}$$

where $\delta_E = 0.015$. Systemic parameters also experience homeostatic relaxation toward un-shocked heterogeneous baseline trajectory states ($\theta$), formulated as:

$$\Delta P_i = \gamma(P_{i,\text{baseline}} - P_i) + \sum_j w_{ji} P_j$$

where $\gamma$ represents the relaxation coefficient governing network recovery speed, and $w_{ji}$ represents the weight of influence from parameter $j$ to parameter $i$.

Pressure is treated as a conditional variable whose operational contribution to PPC effectiveness ($P$) is modeled as a non-monotonic function of its intensity. Rather than acting strictly as a reinforcing or constraining driver, pressure follows an inverted-U response curve: effectiveness drops to zero below an operational floor ($\theta_{\text{floor}} = 0.2$), while values above the threshold are evaluated via a Gaussian response centered at an optimal value ($\mu = 0.5$, $\sigma = 0.28$):

$$P(p) = \begin{cases} 0, & \text{if } p \leq \theta_{\text{floor}} \\ \exp\left(-\frac{1}{2}\left(\frac{p-\mu}{\sigma}\right)^2\right), & \text{if } p > \theta_{\text{floor}} \end{cases}$$

| Parameter | Value | Description |
|---|---|---|

where $p \in [0,1]$ represents the aggregate network pressure at time step $t$. Below $0.2$, actors lack sufficient urgency to collaborate; beyond $0.5$, escalating pressure produces counterproductive friction and trust degradation.

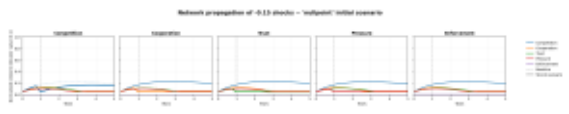

$-0.15$ *shock*

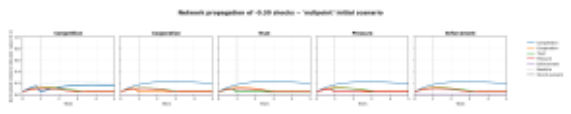

$-0.30$ *shock*

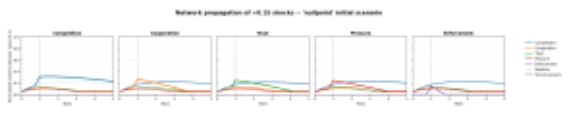

$+0.15$ *shock*

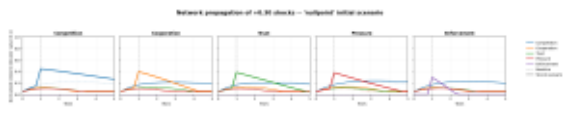

$+0.30$ *shock*

*Time-series response of network indicator values (0–1) over a 5-year period following single-parameter shock interventions for the nullpoint initial scenario. The dashed lines indicate unshocked baseline trajectories; bold solid lines indicate the immediate parameter displacement at Year 1 and subsequent network propagation.*

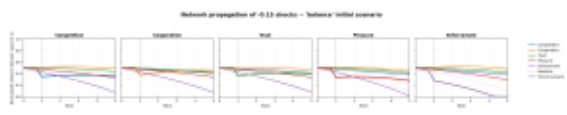

$-0.15$ *shock*

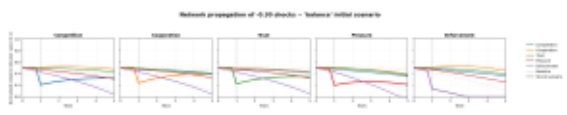

$-0.30$ *shock*

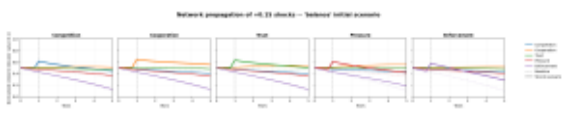

$+0.15$ *shock*

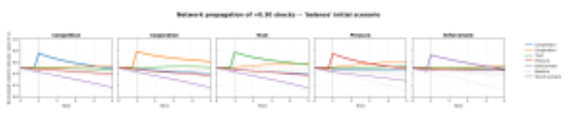

$+0.30$ *shock*

*Time-series response of network indicator values (0–1) over a 5-year period following single-parameter shock interventions for the balance initial scenario. Dashed lines indicate unshocked baseline trajectories; bold solid lines indicate the immediate parameter displacement at Year 1 and subsequent network propagation.*

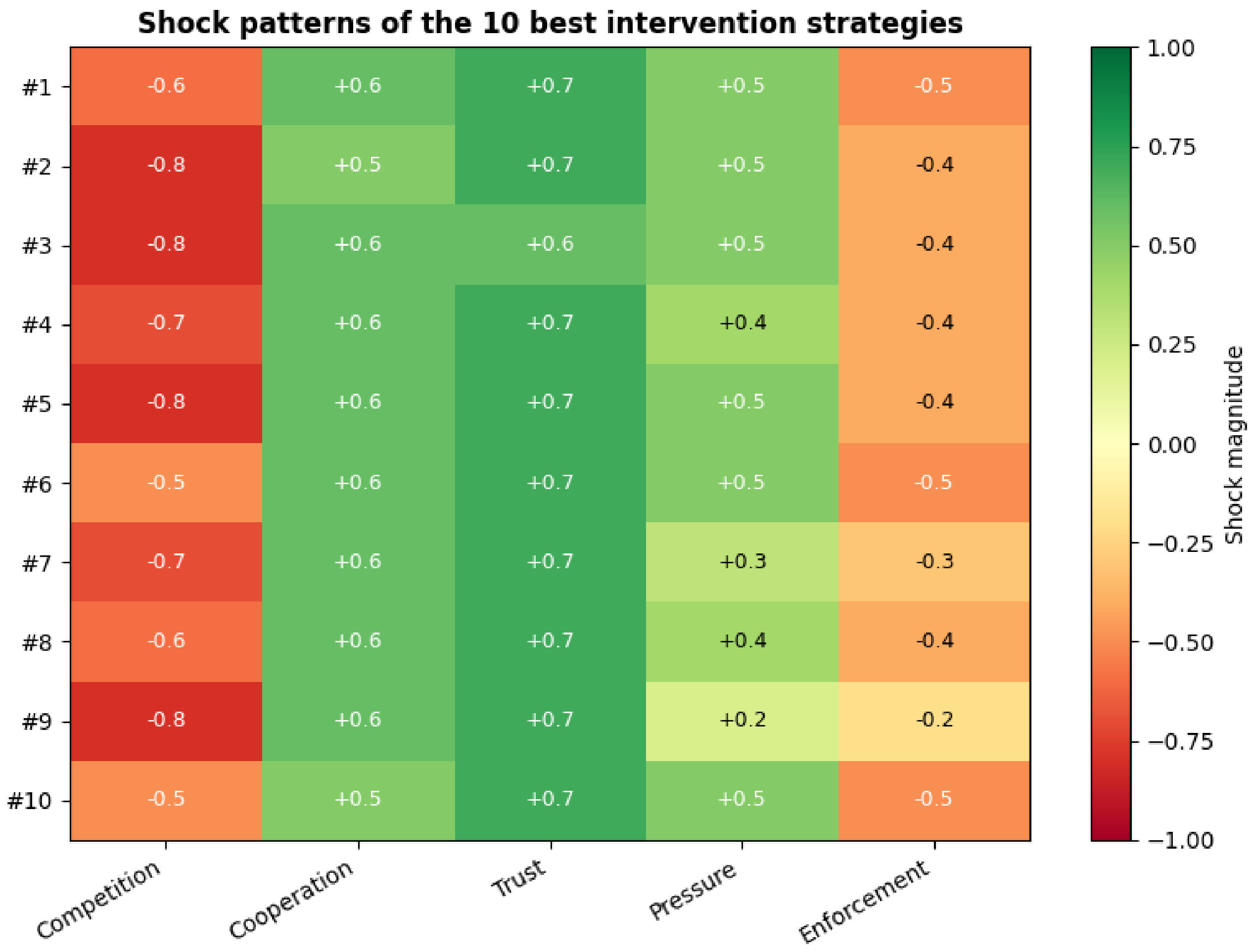

*Top 10 best intervention strategies in regard to effectiveness for heterogeneous scenario with multi-parameter analysis.*

*Top 5 intervention strategies from the exhaustive combinatorial shock analysis (heterogeneous scenario).*

| **Rank** | **Shock pattern** | **PPC effectiveness** | **Δ Effectiveness** |
|---|---|---|---|
| 1 | Competition −0.6, Cooperation +0.6, Trust +0.7, Pressure +0.5, Enforcement −0.5 | 0.568 | 0.246 |
| 2 | Competition −0.8, Cooperation +0.5, Trust +0.7, Pressure +0.5, Enforcement −0.4 | 0.568 | 0.245 |
| 3 | Competition −0.8, Cooperation +0.6, Trust +0.6, Pressure +0.5, Enforcement −0.4 | 0.568 | 0.245 |
| 4 | Competition −0.7, Cooperation +0.6, Trust +0.7, Pressure +0.4, Enforcement −0.4 | 0.567 | 0.245 |
| 5 | Competition −0.8, Cooperation +0.6, Trust +0.7, Pressure +0.5, Enforcement −0.4 | 0.565 | 0.243 |

*Proposed pattern-oriented model checks for future work.*

| Model mechanism | Expected qualitative pattern | Model check |
|---|---|---|
| Trust → cooperation | Higher trust supports cooperation | Positive directional response |
| Cooperation → trust | Sustained cooperation reinforces trust | Positive feedback |
| Competition → cooperation | Higher competition constrains cooperation | Negative directional response |
| Enforcement decay | Enforcement declines without renewal | Multiplicative decay trajectory |
| Parameter → skill | Relational shocks alter actor capacities | Gradual propagation |
| Multi-parameter interaction | Combined changes can differ from isolated changes | Combinatorial analysis |

### 7.0.1 PPC Effectiveness Calculation

The following equation defines the model's output measure of PPC effectiveness; it should be distinguished from the dynamic mechanisms that generate the relational, actor-level, and network states entering the equation.

The PPC effectiveness score is calculated as

$$\mathrm{PPC}_{\mathrm{eff}} = \mathrm{clip}(S \cdot P \cdot (0.25T + 0.25C + 0.20R + 0.15M + 0.15Q) \cdot (0.5 + 0.5F) - H,\ 0,\ 1),$$

where

- $S$ denotes the structural validity gate;
- $P$ denotes the pressure sweet-spot factor;
- $T$ denotes the trust trend gate;
- $C$ denotes the cooperation trend gate;
- $R$ denotes the cooperation–competition gate, requiring cooperation to exceed competition;
- $M$ denotes the mentality gate;
- $Q$ denotes the resource gate;
- $F$ denotes the responsibility term; and
- $H$ denotes the reputation penalty.

The effectiveness score is not a simple average of the five relational indicators. Instead, it represents the degree to which the overall system satisfies a set of institutional, relational, behavioural, and resource conditions considered necessary for successful PPC.

Although the final effectiveness index uses a weighted aggregation of the component gates, the inputs to this aggregation are generated by nonlinear and dynamically evolving processes within the agent-based model, including logistic gating, the nonlinear pressure function, parameter–skill feedback, parameter–parameter interactions, clipping, actor-level modifiers, and network dynamics.

The structural validity gate is binary and ensures that the network contains at least one public actor, one private actor, an actor with decision-making authority, and an actor operating at the operational level. If any of these conditions are not satisfied, PPC effectiveness is reduced to zero. The pressure component follows an inverted-U relationship, reflecting the assumption that both insufficient and excessive pressure reduce collaborative performance. Trust and cooperation are required to remain positive and exhibit an increasing trend over time, while cooperation must exceed competition to favour collaborative rather than adversarial interactions.

The mentality component combines the average levels of risk mentality and punishment mentality, rewarding willingness to cooperate while penalizing excessive aversion to risk. The resource component requires sufficient economic, cultural, and social capital together with an adequate sense of urgency. Formal and informal responsibility jointly increase effectiveness through a multiplicative scaling factor, whereas reputation contributes as a penalty term under the assumption that highly reputation-sensitive actors become increasingly risk-averse over time. After all components have been combined, the resulting effectiveness score is clipped to the interval $[0,1]$, ensuring that the model output remains a normalized measure of overall PPC effectiveness.

**Additional soft-gate implementation**

The model implements several rule-based conditions using a continuous logistic gate rather than a binary indicator. For a margin x, the gate is

$$G(x) = 1 / (1 + \exp(-kx)),$$

where k = 18 controls the steepness of the transition. A positive margin therefore approaches a gate value of 1, while a negative margin approaches 0.

The mentality gate is calculated as

$$M = G(R - P) \times G(R),$$

where R is mean Risk Mentality and P is mean Punishment Mentality.

The first component rewards configurations in which Risk Mentality exceeds Punishment Mentality. The second component requires Risk Mentality itself to be positive. Thus, the mentality contribution approaches one only when both conditions are strongly satisfied.

The trust and cooperation trend gates are calculated analogously, using the estimated temporal slope of the corresponding parameter over the preceding four simulation steps:

T = G(slope(Trust)) × G(Trust)

C = G(slope(Cooperation)) × G(Cooperation).

The cooperation-competition gate is

R = G(Cooperation − Competition_eff),

where Competition_eff includes the applicable private-sector modifier.

These gates are multiplied rather than averaged, so failure of one condition substantially reduces the overall effectiveness score.